\documentclass[prl,twocolumn,floatfix]{revtex4}

\usepackage{amsmath}
\usepackage{amssymb}
\usepackage{graphicx}
\usepackage{dcolumn}
\usepackage{bm}

\def\m#1{\mathrm{#1}}
\def\Eq#1{(\ref{eq:#1})}

\def\epsilon{\varepsilon}
\def\theta{\vartheta}
\def\rho{\varrho}

\begin{document}

%-------------------------------------------------------------------------------

\title{Vapor Pressure of Ionic Liquids}

\author{Markus Bier}
\email{bier@mf.mpg.de}
\author{S.\ Dietrich}
\affiliation
{
   Max-Planck-Institut f\"{u}r Metallforschung, 
   Heisenbergstr. 3,
   70569 Stuttgart,
   Germany, 
   and
   Institut f\"{u}r Theoretische und Angewandte Physik,
   Universit\"{a}t Stuttgart,
   Pfaffenwaldring 57,
   70569 Stuttgart,
   Germany
}

\date{11 November, 2009}

\begin{abstract}
   We argue that the extremely low vapor pressures of room temperature ionic liquids
   near their triple points are due to the \emph{combination} of strong ionic characters
   and of low melting temperatures.
\end{abstract}

\maketitle

%-------------------------------------------------------------------------------

An extremely low vapor pressure (e.g., ca.\ $100\,\m{pPa}$ at $298\,\m{K}$ for 
$\m{[C_4mim][PF_6]}$ \cite{Paulechka2003} compared with $3\,\m{kPa}$ at $298\,\m{K}$ 
for $\m{H_2O}$ \cite{NIST}) is one of the extraordinary properties of room temperature 
ionic liquids (RTILs), i.e., molten salts with melting points below $100^\circ\m{C}$.
As a consequence, RTILs such as $\m{[C_4mim][PF_6]}$ at $298\,\m{K}$ are liquids which
do not evaporate significantly even under ultrahigh vacuum (UHV) conditions (i.e., for a pressure range 
$100\,\m{nPa}\dots100\,\m{pPa}$ \cite{Redhead2003}), which offers the possibility to use 
RTILs, e.g., as substitutes for volatile organic solvents \cite{Wasserscheid2000,Ludwig2007}.
Only a decade ago RTILs were still described as ``non-volatile'' \cite{Wasserscheid2000}, 
but meanwhile direct measurements of their vapor pressures and enthalpies of vaporization at
elevated temperatures have been carried out \cite{Paulechka2005,Zaitsau2006}; even the 
distillation of RTILs \cite{Earle2006} has been achieved. 
Since non-ionic liquids (NILs, such as benzene and water) exhibit triple point pressures $p_3$ above
$1\,\m{Pa}$ (see Tab.~\ref{tab:1}(a)), one might be tempted to attribute the extremely low 
triple point pressures of RTILs exclusively to their ionic character.
However, a comparision of RTILs, which are composed of organic ions, with inorganic fused salts (IFSs),
which are also of ionic character, reveals that the triple point pressures of the latter are above 
$1\,\m{Pa}$ (see Tab.~\ref{tab:1}(c)), such as for NILs.
This rules out that the ionic character is the only reason for the low triple point pressures of RTILs.
We shall show below that it is in fact the \emph{combination} of the melting 
point to occur below \emph{room temperature} and of the \emph{ionic} character of RTILs 
which leads to the observed low triple point pressures.
In other words, any substance with a strong ionic character fulfilling the definition 
of an RTIL inevitably exhibits extremely low vapor pressures near its triple point.

%-------------------------------------------------------------------------------

\begin{figure}[!t]
   \includegraphics{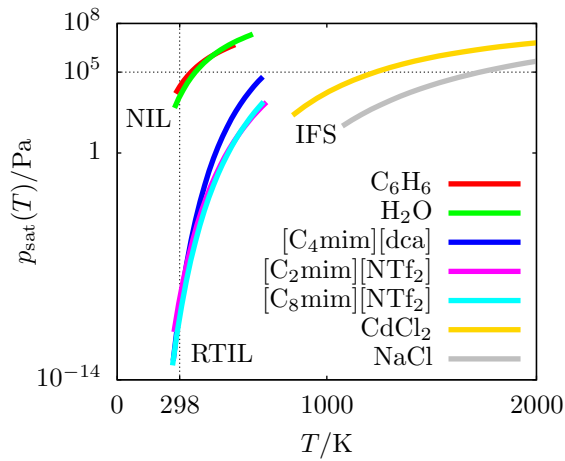}
   \caption{\label{fig:1}Experimental vapor pressures $p_\m{sat}(T)$ at liquid-vapor coexistence of
            non-ionic liquids (NILs), room temperature ionic liquids (RTILs), and inorganic fused salts
            (IFSs) as a function of temperature $T$ for the non-polar liquid benzene ($\m{C_6H_6}$, see Ref.~\cite{NIST}), 
            the hydrogen bond forming liquid water ($\m{H_2O}$, see Ref.~\cite{NIST}), the paradigmatic
            RTILs $\m{[C_4mim][dca]}$, $\m{[C_2mim][NTf_2]}$, and $\m{[C_8mim][NTf_2]}$ 
            (see Refs.~\cite{Emelyanenko2007,Zaitsau2006}), as well as fused cadmium chloride 
            ($\m{CdCl_2}$) and sodium chloride ($\m{NaCl}$) as examples of IFSs (see Ref.~\cite{Barton1956}). 
            At low temperatures all curves terminate at the corresponding triple point temperature 
            $T_3$ (see Tab.~\ref{tab:1}), which is close to the standard melting temperature of that substance.
            At high tempertures the boiling curves for the RTILs terminate at the decomposition 
            temperature $T_d$, whereas the boiling curves of the other liquids end at their critical
            points (see Tab.~\ref{tab:1}). Room temperature $T_0=298\,\m{K}$ and ambient pressure $p_0=10^5\,\m{Pa}$ are
            indicated.}
\end{figure}

\begin{table}[!t]
   \begin{tabular}{|l|c|c|c|c|l|}
      \hline
      (a) NIL      & $T_3 / \m{K}$ & $p_3 / \m{Pa}$ & $T_b / \m{K}$ & $T_c / \m{K}$ & Refs.       \\
      \hline
      $\m{C_6H_6}$ & $278.7$       & $4799$         & $353.2$       & $562.1$       & \cite{NIST} \\
      $\m{H_2O}$   & $273.2$       & $611.7$        & $373.1$       & $647.1$       & \cite{NIST} \\
      \hline
   \end{tabular} 
 
   \bigskip

   \begin{tabular}{|l|c|c|c|c|l|}
      \hline
      (b) RTIL              & $T_3 / \m{K}$ & $p_3 / \m{Pa}$   & $T_d / \m{K}$ & $T^\m{extr}_b / \m{K}$ & Refs.                                       \\
      \hline
      $\m{[C_4mim][dca]}$   & $267$         & $1.5\times10^{-13}$ & $695$         & $719$                & \cite{Fredlake2004,Emelyanenko2007}         \\
      $\m{[C_2mim][NTf_2]}$ & $271$         & $8.9\times10^{-12}$ & $712$         & $906$                & \cite{Zaitsau2006,Paulechka2007,Tokuda2005} \\
      $\m{[C_8mim][NTf_2]}$ & $264$         & $7.8\times10^{-14}$ & $698$         & $857$                & \cite{Zaitsau2006,Paulechka2007,Tokuda2005} \\
      \hline
   \end{tabular} 

   \bigskip

   \begin{tabular}{|l|c|c|c|c|l|}
      \hline
      (c) IFS      & $T_3 / \m{K}$ & $p_3 / \m{Pa}$ & $T_b / \m{K}$ & $T_c / \m{K}$ & Refs.                                    \\
      \hline
      $\m{CdCl_2}$ & $837$         & $214$          & $1233$        & ?             & \cite{Lide1998,Barton1956}               \\
      $\m{NaCl}$   & $1074$        & $46$           & $1738$        & $>3400$       & \cite{Lide1998,Barton1956,McGonigal1963} \\
      \hline
   \end{tabular} 

   \caption{\label{tab:1}Experimental data for characteristic temperatures of (a) non-ionic liquids 
            (NILs), (b) room temperature ionic liquids (RTILs), and (c) inorganic fused 
            salts (IFSs) corresponding to the substances discussed in Fig.~\ref{fig:1}.
            $T_3$ and $p_3$ denote the temperature and the pressure, respectively, at the triple point, 
            $T_c$ is the critical temperature, and $T_d$ denotes the temperature for the onset of
            decomposition of an RTIL \cite{Fredlake2004,Tokuda2005}.
            $T_b$ denotes the standard boiling temperature at ambient pressure $p_0=10^5\,\m{Pa}$ for NILs and IFSs, whereas 
            the standard boiling temperatures $T^\m{extr}_b$ for RTILs are estimated by extrapolation \cite{Rebelo2005}
            because boiling of RTILs is preempted by decomposition.} 
\end{table}

Figure~\ref{fig:1} displays the experimental vapor pressures $p_\m{sat}(T)$ for liquid-vapor coexistence
at temperature $T$ for the non-polar liquid benzene ($\m{C_6H_6}$, see Ref.~\cite{NIST}), the hydrogen bond 
forming liquid water ($\m{H_2O}$, see Ref.~\cite{NIST}), the paradigmatic
RTILs $\m{[C_4mim][dca]}$, $\m{[C_2mim][NTf_2]}$, and $\m{[C_8mim][NTf_2]}$ 
(see Refs.~\cite{Emelyanenko2007,Zaitsau2006}), as well as fused cadmium chloride 
($\m{CdCl_2}$) and sodium chloride ($\m{NaCl}$) as representatives of IFSs (see Ref.~\cite{Barton1956}). 
At low temperatures the boiling curves terminate at the triple point temperature 
$T_3$ (see Tab.~\ref{tab:1} and Refs.~\cite{NIST,Fredlake2004,Paulechka2007,Lide1998}), which is close the standard 
melting temperature of the corresponding substance because the melting curve is very steep.
At high temperatures the boiling curves of the NILs and the IFSs terminate at their critical 
temperatures $T_c$ (see Tabs.~\ref{tab:1}(a) and (c) and Refs.~\cite{NIST,McGonigal1963}), whereas RTILs decompose at a
substance specific decomposition temperature $T_d$ (see Tab.~\ref{tab:1}(b) and 
Refs.~\cite{Wasserscheid2000,Fredlake2004,Tokuda2005}).
As it is apparent from Fig.~\ref{fig:1}, RTILs do not boil at ambient pressure $p_0=10^5\,\m{Pa}$ because boiling is
preempted by decomposition; consequently Tab.~\ref{tab:1}(b) displays only extrapolated standard
boiling temperatures $T_b^\m{extr}$ for RTILs.

%-------------------------------------------------------------------------------

In order to understand the position of the boiling curves of RTILs in Fig.~\ref{fig:1},
we note that with respect to the strength of the particle-particle interaction, RTILs lie in 
between NILs, which interact via relatively weak dispersion forces and possibly 
hydrogen bonds, and IFSs, which interact predominantly via strong Coulomb forces.
Due to the larger size of the RTIL ions and a possible delocalization of the charge
their interaction is, however, weaker than that of IFS ions.
Hence, ignoring for the time being the decomposition of RTILs at $T_d$, the molar enthalpies 
of vaporization $\Delta_\m{vap}H(p)>0$ at pressure $p$ are expected to be ordered as 
$\Delta_\m{vap}H^\m{NIL}(p) < \Delta_\m{vap}H^\m{RTIL}(p) < \Delta_\m{vap}H^\m{IFS}(p)$.
On the other hand, in the spirit of Trouton's rule \cite{Atkins1998}, the molar entropies 
of vaporization $\Delta_\m{vap}S(p)$ at pressure $p$ are expected to depend only weakly on 
the kind of substance, because their values are dominated by the translational and rotational degrees of 
freedom whereas vibrational and electronic modes and the structural arrangements contribute only 
as small corrections \cite{Vincett1978}.
Data for organic and inorganic liquids tabulated in Refs.~\cite{Lide1998,Janz1967}
suggest a Trouton-like rule $\Delta_\m{vap}S(p_0)\approx (95\pm15)\,\m{J/mol}$ at ambient pressure 
$p_0=10^5\,\m{Pa}$.
According to $\Delta_\m{vap}H(p)=T_b(p)\Delta_\m{vap}S(p)$ \cite{Atkins1998} with $T_b(p)$ 
denoting the boiling temperature at pressure $p$ one expects the relation
$T_b^\m{NIL}(p) < T_b^\m{RTIL}(p) < T_b^\m{IFS}(p)$, which is indeed consistent with the
experimental findings for NILs and IFSs \cite{Janz1967,Lide1998,NIST} and the extrapolations for RTILs to 
ambient pressure \cite{Rebelo2005} (see also Fig.~\ref{fig:1} and Tab.~\ref{tab:1}).
Away from the critical point $\Delta_\m{vap}H(p)$ and $\Delta_\m{vap}S(p)$ depend only weakly on 
$p$ \cite{Lide1998,NIST}, such that we can approximate $\Delta_\m{vap}H(p)\approx\Delta_\m{vap}H(p_0)$ 
and $\Delta_\m{vap}S(p)\approx\Delta_\m{vap}S(p_0)$ for a certain reference pressure $p_0$ such as the 
ambient pressure.
Within this approximation the Clausius-Clapeyron equation \cite{Atkins1998} allows one to estimate the
vapor pressure $p_\m{sat}(T)$ for liquid-vapor coexistence at temperature $T$:
\begin{equation}
   p_\m{sat}(T) \approx 
   p_0\exp\Big(-\frac{\Delta_\m{vap}H(p_0)}{RT} + \frac{\Delta_\m{vap}S(p_0)}{R}\Big).
   \label{eq:ClauClap}
\end{equation}
According to the above reasoning concerning $\Delta_\m{vap}H$ and $\Delta_\m{vap}S$ one infers
the relation 
\begin{equation}
   p_\m{sat}^\m{NIL}(T) \gg p_\m{sat}^\m{RTIL}(T) \gg p_\m{sat}^\m{IFS}(T).
   \label{eq:psat}
\end{equation} 
Actually, liquid-vapor coexistence at $p_\m{sat}(T)$ occurs only in the temperature ranges 
$T_3 \leq T \leq T_c$ for NILs and IFSs and $T_3 \leq T \leq T_d$ for RTILs (see Fig.~\ref{fig:1}).

%-------------------------------------------------------------------------------

Equation~\Eq{psat} follows from general considerations concerning the strength of the 
particle-particle interaction and the entropy of vaporization.
According to these simple arguments it is indeed the strong ionic character which leads to a downshift
of the boiling curves $p_\m{sat}(T)$ of RTILs relative to those of NILs.
However, the reason for IFSs having not an even lower triple point pressure than RTILs is 
the large difference in the triple point temperatures ($T_3^\m{RTIL} < T_3^\m{IFS}$, see Tab.~\ref{tab:1})
induced by a large difference in standard melting temperatures.
The mechanism for leading to the low standard melting temperatures of
RTILs has been explained in terms of a frustrated crystallization due to asymmetric ion 
shapes, charge delocalization, packing inefficiency, and conformational degeneracy 
\cite{Seddon1997,Larsen2000,Holbrey2003,Plechkova2008}.
Hence the extremely low vapor pressues of RTILs near their triple points can be 
understood on very general grounds based on both a strong ionic character \emph{and}
low melting temperatures; the conclusions are independent of substance specific 
properties which explains why this phenomenon is a common feature of RTILs.

%-------------------------------------------------------------------------------

In summary, we have shown that near its triple point the vapor pressure of a room 
temperature ionic liquid of strong ionic character is very small, because it depends 
exponentially on the ratio of a large enthalpy of vaporization --- which 
is almost as large as that of inorganic salts --- and a small thermal energy near the
triple point, which is as small as that of non-ionic liquids.
According to $p_\m{sat}(T)\sim\exp(-\Delta_\m{vap}H/(RT))$, where the prefactor is
approximately independent of the kind of substance, an increase of $\Delta_\m{vap}H$, reflecting
the ionic character of room temperature ionic liquids relative to non-ionic liquids,
leads to a downshift of $p_\m{sat}(T)$.
For room temperature ionic liquids these low vapor pressures are physically accessible
due to their low triple points, induced by their low melting temperature --- which is
part of the definition of room temperature ionic liquids (see Fig.~\ref{fig:1}).
The even stronger ionic character of inorganic fused salts would in principle lead to
even lower vapor pressures; however, these cannot be reached for their liquid state
because they are preempted by a significantly higher freezing and thus triple point
temperature (see Fig.~\ref{fig:1}). 

%-------------------------------------------------------------------------------

%-------------------------------------------------------------------------------


\begin{thebibliography}{00}
   \bibitem{Paulechka2003}
      Y.\ U.\ Paulechka, G.\ J.\ Kabo, A.\ V.\ Blokhin, O.\ A.\ Vydrov, J.\ W.\ Magee, and M.\ Frenkel,
      J.\ Chem.\ Eng.\ Data \textbf{48}, 457 (2003).
   \bibitem{NIST}
      E.\ W.\ Lemmon, M.\ O.\ McLinden, and D.\ G.\ Friend,
      \textit{Thermophysical Properties of Fluid Systems}, in NIST Chemistry WebBook,
      NIST Standard Reference Database Number 69,
      edited by P.\ J.\ Linstrom and W.\ G.\ Mallard,
      \texttt{http://webbook.nist.gov}
   \bibitem{Redhead2003}
      P.\ A.\ Redhead,
      J.\ Vac.\ Sci.\ Technol.\ A \textbf{21}, S12 (2003).
   \bibitem{Wasserscheid2000}
      P.\ Wasserscheid and W.\ Keim,
      Angew.\ Chem.\ Int.\ Ed.\ \textbf{39}, 3773 (2000).
   \bibitem{Ludwig2007}
      R.\ Ludwig and U.\ Kragl,
      Angew.\ Chem.\ Int.\ Ed.\ \textbf{46}, 6582 (2007).
   \bibitem{Paulechka2005}
      Y.\ U.\ Paulechka, Dz.\ H.\ Zaitsau, G.\ J.\ Kabo, and A.\ A.\ Stechan,
      Thermochim.\ Acta \textbf{439}, 158 (2005).
   \bibitem{Zaitsau2006}
      Dz.\ Zaitsau, G.\ J.\ Kabo, A.\ A.\ Stechan, Y.\ U.\ Paulechka, 
      A.\ Tschersich, S.\ P.\ Verevkin, and A.\ Heintz,
      J.\ Phys.\ Chem.\ A \textbf{110}, 7303 (2006).
   \bibitem{Earle2006}
      M.\ J.\ Earle, J.\ M.\ S.\ S.\ Esperan\c{c}a, M.\ A.\ Gilea, 
      J.\ N.\ C.\ Lopes, L.\ P.\ N.\ Rebelo, J.\ W.\ Magee, K.\ R.\ Seddon, 
      and J.\ A.\ Widegren,
      Nature \textbf{439}, 831 (2005).  
   \bibitem{Emelyanenko2007}
      V.\ N.\ Emel'yanenko, S.\ P.\ Verevkin, and A.\ Heintz,
      J.\ Am.\ Chem.\ Soc.\ \textbf{129}, 3930 (2007).
   \bibitem{Barton1956}
      J.\ L.\ Barton and H.\ Bloom,
      J.\ Phys.\ Chem.\ \textbf{60}, 1413 (1956).
   \bibitem{Fredlake2004}
      C.\ P.\ Fredlake, J.\ M.\ Crosthwaite, D.\ G.\ Hert, S.\ N.\ V.\ K.\ Aki, 
      and J.\ F.\ Brennecke,
      J.\ Chem.\ Eng.\ Data \textbf{49}, 954 (2004).
   \bibitem{Lide1998}
      D.\ R.\ Lide (Ed.),
      \textit{CRC Handbook of Chemistry and Physics}
      (CRC Press, Boca Raton, 1998).
   \bibitem{Paulechka2007}
      Y.\ U.\ Paulechka, A.\ V.\ Blokhin, G.\ J.\ Kabo, and A.\ A.\ Stechan,
      J.\ Chem.\ Thermodyn.\ \textbf{39}, 866 (2007).
   \bibitem{McGonigal1963} 
      P.\ J.\ McGonigal,
      J.\ Phys.\ Chem.\ \textbf{67}, 1931 (1963).
   \bibitem{Tokuda2005}
      H.\ Tokuda, K.\ Hayamizu, K.\ Ishii, M.\ A.\ Bin Hasan Susan, 
      and M.\ Watanabe,
      J.\ Phys.\ Chem.\ B \textbf{109}, 6103 (2005).
   \bibitem{Atkins1998}
      P.\ W.\ Atkins,
      \textit{Physical chemistry}, 6th Ed. 
      (Oxford University Press, Oxford, 1998).
   \bibitem{Vincett1978}
      P.\ S.\ Vincett,
      J.\ Phys.\ Chem.\ \textbf{82}, 2797 (1978).
   \bibitem{Janz1967}
      G.\ J.\ Janz,
      \textit{Molten Salts Handbook}
      (Academic Press, New York, 1967).
   \bibitem{Rebelo2005}
      L.\ P.\ N.\ Rebelo, J.\ N.\ C.\ Lopes, J.\ M.\ S.\ S.\ Esperan\c{c}a, 
      and E.\ Filipe,
      J.\ Phys.\ Chem.\ B \textbf{109}, 6040 (2005).
   \bibitem{Seddon1997}
      K.\ R.\ Seddon,
      J.\ Chem.\ Tech.\ Biotechnol.\ \textbf{68}, 351 (1997).
   \bibitem{Larsen2000}
      A.\ S.\ Larsen, J.\ D.\ Holbrey, F.\ S.\ Tham, and C.\ A.\ Reed,
      J.\ Am.\ Chem.\ Soc.\ \textbf{122}, 7264 (2000).
   \bibitem{Holbrey2003}
      J.\ D.\ Holbrey, W.\ M.\ Reichert, M.\ Nieuwenhuyzen, S.\ Johnston, 
      K.\ R.\ Seddon, and R.\ D.\ Rogers,
      Chem.\ Commun.\ \textbf{2003}, 1636.
   \bibitem{Plechkova2008}
      N.\ V.\ Plechkova and K.\ R.\ Seddon,
      Chem.\ Soc.\ Rev.\ \textbf{37}, 123 (2008).
\end{thebibliography}
\end{document}